\documentclass{article}
\usepackage{spconf,amsmath,amssymb,graphicx}
\usepackage[caption=false,font=footnotesize]{subfig}
\usepackage{booktabs}
\usepackage{siunitx}
\usepackage{tikz}
\usetikzlibrary{arrows.meta,positioning,fit}
\tikzset{
  >={Latex[length=2.2mm,width=1.8mm]},
  block/.style={draw,rounded corners=2pt,very thick,align=center,
                inner xsep=3mm,inner ysep=2mm,minimum height=9mm},
  group/.style={draw,dashed,rounded corners=2pt,inner sep=3mm}
}
\usepackage[hidelinks]{hyperref} % תמיד אחרון

\newcommand{\startrefsafterfourpages}{%
  \clearpage                         % לסגור פְּלוֹאוֹטִים ולעבור לעמוד הבא
  \ifnum\value{page}<4 \null\thispagestyle{empty}\newpage \fi
  \ifnum\value{page}<4 \null\thispagestyle{empty}\newpage \fi
  \ifnum\value{page}<4 \null\thispagestyle{empty}\newpage \fi
  \newpage
}

\usepackage{xcolor}
\usepackage{graphicx}
\graphicspath{{figs/}}

\title{An Efficient Neural Network for Modeling Human Auditory Neurograms for Speech}

\name{Eylon Zohar$^{1}$ \qquad Israel Nelken$^{2}$ \qquad Boaz Rafaely$^{1}$}
\address{%
$^{1}$ School of Electrical and Computer Engineering, Ben-Gurion University of the Negev, Israel\\
$^{2}$ Department of Neurobiology, The Hebrew University of Jerusalem, Israel\\
\texttt{eylonz1997@gmail.com} \quad
\texttt{israel.nelken@mail.huji.ac.il} \quad
\texttt{br@bgu.ac.il}
}

\begin{document}
\ninept
\maketitle

\begin{abstract}
Classical auditory-periphery models, exemplified by Bruce et al., 2018, provide high-fidelity simulations but are stochastic and computationally demanding, limiting large-scale experimentation and low-latency use. Prior neural encoders approximate aspects of the periphery; however, few are explicitly trained to reproduce the deterministic, rate-domain neurogram , hindering like-for-like evaluation.
We present a compact convolutional encoder that approximates the Bruce mean-rate pathway and maps audio to a multi-frequency neurogram. We deliberately omit stochastic spiking effects and focus on a deterministic mapping (identical outputs for identical inputs). Using a computationally efficient design, the encoder achieves close correspondence to the reference while significantly reducing computation, enabling efficient modeling and front-end processing for auditory neuroscience and audio signal processing applications.
\end{abstract}


\section{Introduction}
Modeling how the auditory system transforms sound into neural representations is central to both neuroscience and audio signal processing. A key representation is the neurogram -- a biologically inspired time--frequency map of activity across auditory-nerve fibers (ANFs) as a function of characteristic frequency (CF) and time -- used in auditory neuroscience, brain--computer interfaces (BCIs), hearing prostheses, and neuromorphic audio. Generating such simulated neurograms from waveforms typically involves nonlinear stages that emulate cochlear filtering, inner hair cell (IHC) transduction, and synaptic adaptation in the auditory periphery~\cite{bruce2018,zilany2009,zilany2014,amtoolbox_acta2022}.

Classical auditory--periphery models such as Zilany et~al.~\cite{zilany2014}, Bruce et~al.~\cite{bruce2018}, and the AMT implementation~\cite{amtoolbox_acta2022} provide high-fidelity simulated neurograms. At the same time, stochastic synapses and computational demand in these references can complicate large-scale experiments and integration with gradient-based pipelines that require stable supervision~\cite{bruce2018,zilany2009,zilany2014,amtoolbox_acta2022}. A fully differentiable encoder that yields reproducible, rate-domain targets would facilitate rigorous evaluation and seamless use as a front end in modern learning systems.

Neural surrogates have therefore been proposed. In particular, CoNNear learns convolutional approximations for cochlear, inner hair cell (IHC), and auditory-nerve fiber (ANF) processing by driving analytical reference models with speech and optimizing an L1 (mean absolute error, MAE) objective to match stage-specific outputs~\cite{baby2021connear,verhulst2021}. Critically, CoNNear’s validation is not a quantitative, speech-based comparison to the analytical reference: it primarily examines cochlear and auditory-nerve properties using pure-tone and click probes~\cite{baby2021connear,verhulst2021}, whereas continuous speech is not the main evaluation target. Related focus on cochlear mechanics only or produce spike-based representations~\cite{pan2019,verma2019}. As a result, the literature does not report a speech-centric quantitative comparison between learned surrogates and an analytical auditory-nerve reference.

In this work, we present a fully differentiable auditory encoder that maps raw audio to simulated, multi--CF neurograms while emulating the deterministic (noise-free) rate-domain pathway of the Bruce auditory-nerve model via its AMT implementation~\cite{bruce2018,amtoolbox_acta2022}. Our evaluation is speech-first: we quantify fidelity to the Bruce reference on continuous speech using normalized mean-squared error (NMSE), Pearson correlation, and signal-to-noise ratio (SNR) (precise definitions appear in Sec.~\ref{sec:exp}). In addition, we motivate and employ a bandwise design that partitions CFs into three groups to stabilize optimization across disparate dynamic ranges.

\section{Auditory Model}\label{sec:bruce}

We use the auditory-nerve model of Bruce et al.~\cite{bruce2018} as the physiological reference,
via its \texttt{bruce2018} implementation in the Auditory Modeling Toolbox (AMT)~\cite{amtoolbox_acta2022,amtoolbox2022}.
The model comprises cochlear filtering, IHC transduction, and synaptic dynamics.
The original release includes a stochastic synaptic-release mechanism; in this study we focus on the
\emph{deterministic} mean-rate pathway to obtain reproducible simulated neurograms.
Here, \emph{deterministic} means that identical inputs yield identical outputs. Stochastic effects are
treated as small at the resolutions considered and are neglected as a first-order approximation.

\graphicspath{{figures/}} % אם הקובץ יושב בתיקייה figures/

% בגוף המסמך – שים את הבלוק הזה בסוף עמוד 1 (ראו "איפה לשים" למטה)
\begin{figure*}[!t]
  \centering
  % רוחב שני עמודות; גובה מוגבל כדי שלא "יאכל" חצי עמוד
  \includegraphics[width=\textwidth,height=0.28\textheight,keepaspectratio]{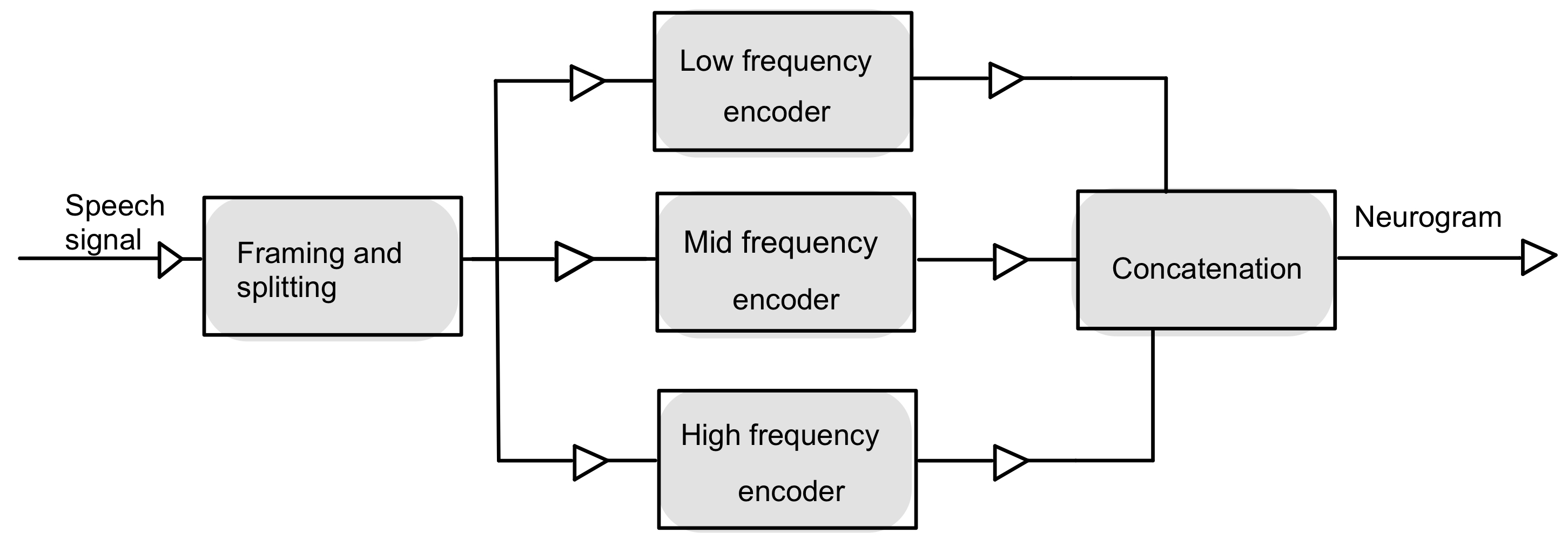}
  \caption{{Block diagram of the inference path, showing the framing and frequency splitting block, followed by the three encoders and the concatenation of their  output to a single neurogram.}}
  \label{fig:proposed_encoder_blockdiagram}
\end{figure*}
% אם צריך לצמצם רווחים אחרי האיור:
\vspace{-2mm}

% ===================== Section =====================
\section{Proposed Neural Encoder}\label{sec:encoder}

\subsection{Overview}\label{sec:overview}
Figure~\ref{fig:proposed_encoder_blockdiagram} illustrates the inference path of the proposed encoder. A clean speech waveform was first preprocessed and split into three frequency bands. These band-specific signals then served as inputs to three independent convolutional encoders (one per band). The resulting sub-neurograms were concatenated along the CF axis to produce the final deterministic rate-domain neurogram.

\subsection{Preprocessing and splitting}\label{sec:arch}
A clean speech waveform was framed into \SI{150}{ms} windows with \SI{100}{ms} overlap (\SI{50}{ms} hop size). Each frame was mapped onto an equivalent rectangular bandwidth (ERB) -spaced CF grid and split into three disjoint groups: low (CF 0–16), mid (CF 17–28), and high (CF 29–32). These band signals served as inputs to the band-specific encoders in Sec.~\ref{sec:freq_enc}.

\subsection{Frequency-band specific encoders}\label{sec:freq_enc}
We use three independent encoders for disjoint CF groups: low (CF 0--16; 17 ch), mid (CF 17--28; 12 ch), and high (CF 29--32; 4 ch). Each encoder receives $X_b \in \mathbb{R}^{|F_b|\times 15000}$ (CF$\times$time; 150\,ms at 100\,kHz) and predicts $\hat N_b \in \mathbb{R}^{|F_b|\times 5000}$ for the last 50\,ms; concatenation along CF yields a $33\times 5000$ neurogram. The encoder takes the form of a compact TCN-style design with multi-dilation temporal CNNs as in Conv-TasNet [10]. Training leverages multi-resolution spectral and envelope constraints in line with recent audio models (Parallel WaveGAN~\cite{yamamoto2020parallelwavegan}, HiFi-GAN~\cite{kong2020hifigan}, CoNNear~\cite{baby2021connear}).

\paragraph*{Per-band architecture}
The front-end comprises two causal Conv1D layers. Kernel sizes correspond to the three bands: low $=21$, mid $=11$, high $=7$. Each convolution is followed by GroupNorm and LeakyReLU (the first activation uses GELU).

The encoder stack contains $3$ spectral--temporal blocks. Each block fuses a spectral $1{\times}1$ branch with three temporal branches at dilations $\{1,3,9\}$ and kernel size $3$, followed by average pooling by a factor of $2$ per block. Over three blocks this yields an overall temporal downsampling of $2\times2\times2=8$.

A lightweight attention module operates at the downsampled rate to stabilize longer-range dependencies.

The decoder consists of three upsampling stages (linear upsampling by a factor of $2$ per stage; overall $\times 8$) with skip connections from the matching encoder resolutions.

A final $1{\times}1$ head projects to $|F_b|$ CF channels and is realized with a ReLU to enforce nonnegativity.

All convolutions are causal, with GroupNorm+LeakyReLU applied throughout (GELU in the input stack). Hidden widths inside the blocks are $\{64,128,256\}$ channels (from shallow to deep). The top encoder width per band is: low $=512$ channels, mid $=384$ channels, high $=320$ channels.

\subsection{Concatenation and encoding output}\label{sec:concat}
Outputs from the three encoders were concatenated along the CF axis to form the complete neurogram. The resulting representation comprised 33 characteristic-frequency (CF) channels and 5{,}000 time samples, corresponding to the last \SI{50}{ms} of the frame at \SI{100}{kHz}. The three band outputs were produced in parallel from the same input frame and then aligned without overlap or interpolation.

\subsection{Training}\label{sec:training}

The three frequency-band specific encoders were trained jointly with a single optimizer using a weighted sum of three complementary objectives: a pointwise time-domain term, a multi-resolution STFT magnitude term, and a band-averaged envelope term. This follows common practice in neural audio modeling, where temporal fidelity, spectral structure, and slow-varying dynamics are optimized together \cite{pascual2017segan, rethage2018wavenetdenoise, luo2019convtasnet, yamamoto2020parallelwavegan, kong2020hifigan, baby2021connear, bernstein2013stm, vuong2022modloss, taal2011stoi, jorgensen2013sepsm}.

Formally,
\[
\mathcal{L}_{\mathrm{joint}}
=\sum_{b=1}^{3}\Big(\alpha\,\mathcal{L}^{(b)}_{\mathrm{time}}
+\beta\,\mathcal{L}^{(b)}_{\mathrm{spec}}
+\gamma\,\mathcal{L}^{(b)}_{\mathrm{env}}\Big),
\]
with fixed weights \(\alpha{=}0.5,\ \beta{=}0.3,\ \gamma{=}0.2\) selected on a validation set. Here \(b\in\{1,2,3\}\) indexes the low/mid/high bands.

\textbf{Time-domain loss. The time-domain loss is defined as.}
\[
\mathcal{L}^{(b)}_{\mathrm{time}}
=\frac{1}{|F_b|\,T}\sum_{f\in F_b}\sum_{t=1}^{T}\big(N(f,t)-\hat N(f,t)\big)^2,
\]
where \(T{=}5000\) samples correspond to the last 50\,ms at 100\,kHz, \(N(f,t)\) is the Bruce-derived target neurogram, and \(\hat N(f,t)\) is the estimate. Time-domain objectives of this family (L1/L2 on the waveform or SI-SNR) are standard in end-to-end speech modeling \cite{pascual2017segan, rethage2018wavenetdenoise, luo2019convtasnet}.

\textbf{Multi-resolution STFT loss.}
To capture short- and mid-term spectral envelopes, we compute an STFT \emph{along time} for each CF channel independently. Let \(\mathcal{M}\) be the set of configurations \(m\) with window \(w_m\in\{64,128,256,512,1024\}\) samples, hop size \(h_m{=}w_m/4\), and FFT size \(K_m{=}w_m\); \(U_m\) is the number of TF bins for configuration \(m\), and \(\varepsilon{=}10^{-8}\). Define \(\phi_m(x)\triangleq\log(|S_m(x)|+\varepsilon)\), where \(S_m(\cdot)\) is the complex STFT. The loss is
\[
\mathcal{L}^{(b)}_{\mathrm{spec}}
=\frac{1}{|F_b|\,|\mathcal{M}|}
\sum_{f\in F_b}\sum_{m\in\mathcal{M}}
\frac{1}{U_m}\,\big\|\,\phi_m(N(f,:))-\phi_m(\hat N(f,:))\,\big\|_{2}^{2}.
\]
Each configuration is normalized by its own \(U_m\) and then averaged, so identical frequency sampling across \(\mathcal{M}\) is not assumed. Multi-resolution STFT objectives of this form are widely used to stabilize training and match perceptually relevant spectra \cite{yamamoto2020parallelwavegan, kong2020hifigan}.

\textbf{Envelope loss (band-averaged).}
To emphasize slow dynamics at the band level while avoiding cross-band coupling, we form a \emph{band-averaged envelope} by averaging across CF channels within each band and comparing the resulting 1-D traces via MSE:
\[
E_b(t)=\frac{1}{|F_b|}\sum_{f\in F_b} N(f,t),\ \ 
\hat E_b(t)=\frac{1}{|F_b|}\sum_{f\in F_b} \hat N(f,t),
\]
\[
\mathcal{L}^{(b)}_{\mathrm{env}}
=\frac{1}{T}\sum_{t=1}^{T}\big(E_b(t)-\hat E_b(t)\big)^2.
\]
The motivation for including an envelope term is grounded in auditory-inspired neural modeling \cite{baby2021connear}, modulation-based predictors of intelligibility \cite{bernstein2013stm, taal2011stoi, jorgensen2013sepsm}, and recent proposals of explicit modulation-domain training losses \cite{vuong2022modloss}. These works demonstrate that accurate modeling of envelopes and modulation spectra improves perceptual fidelity and speech intelligibility predictions.

Because the bands have disjoint parameters and no cross-band layers, summing per-band objectives is effectively equivalent—up to optimizer dynamics—to optimizing each band separately. Accordingly, we also report an ablation in which each encoder was trained with its own optimizer using only the temporal term,
\[
\mathcal{L}^{(b)}_{\mathrm{indep}}=\mathcal{L}^{(b)}_{\mathrm{time}},
\]
providing a fully decoupled baseline. In practice, the multi-term joint formulation yielded more stable optimization without introducing cross-band coupling.

\section{Experimental Setup and Methodology}\label{sec:exp}
We validate the proposed encoder on clean speech using a simple and reproducible protocol. This section describes the dataset and target generation, the segmentation used for temporal context, the training procedure and reproducibility settings, the evaluation metrics with their exact definitions presented individually, and the runtime measure.

\subsection{Dataset and Ground-Truth Generation}\label{sec:data}
We used the single-speaker source tracks from the WSJ0-2mix recipe \cite{hershey2015deepclustering}, restricted to clean, single-channel sources (no mixtures). The audio originated from WSJ0 (LDC93S6A; licensed) at \SI{16}{\kilo\hertz}. Ground-truth neurograms $N(f,t)$ were generated with the deterministic (non-stochastic) rate pathway of AMT’s \texttt{bruce2018} \cite{amtoolbox_acta2022,amtoolbox2022} on an ERB-spaced CF grid. Unless stated otherwise, training losses and reported metrics were computed for CF channels $0$–$32$ (\emph{33 channels spanning ERB-spaced center frequencies from approximately \SI{125}{\hertz} up to \SI{8}{\kilo\hertz}}), which in this configuration cover the speech band up to \SI{8}{\kilo\hertz}.

The dataset was split at the \emph{file level} into \SI{80}{\percent}/\SI{10}{\percent}/\SI{10}{\percent} train/validation/test, so that no audio file contributed segments to more than one split. From these files we constructed $120{,}000$ examples, each with a \SI{50}{\milli\second} target window, totaling \SI{100}{\minute} of labeled targets ($96{,}000/12{,}000/12{,}000$ for train/validation/test, respectively). WSJ0 (LDC93S6A) is a licensed LDC corpus; we provide scripts to regenerate the segments from licensed WSJ0 audio upon request \cite{LDC93S6A,hershey2015deepclustering}.

\subsection{Segmentation and Context}\label{sec:segmentation}
Each supervised example consisted of \SI{150}{ms} of input context together with a \SI{50}{ms} output window corresponding to the final part of the frame. The leading \SI{100}{ms} supplied causal history for cochlear and inner–hair–cell filtering and for the deterministic adaptation stages of the Bruce pathway. Consecutive examples were extracted with a \SI{50}{ms} hop, so that an utterance of length $L$ seconds yielded $\lfloor L/0.05\rfloor$ examples. Zero padding was applied only at the utterance onset to provide the initial \SI{100}{ms} of context; no padding was added at the end. Waveforms were mean-removed and scaled to unit standard deviation \emph{per recording} prior to segmentation.

Using \SI{150}{ms} of context covered typical group delays and short-to-mid adaptation time constants, which reduced boundary artifacts—systematic discrepancies at the beginning of the \SI{50}{ms} target window caused by insufficient causal history—and mitigated edge errors at frame boundaries during streaming. Empirically, contexts shorter than \SI{100}{ms} increased these boundary effects, whereas \SI{150}{ms} yielded stable targets across speakers while maintaining low latency for streaming.

\subsection{Training Procedure}\label{sec:trainproc}
Models are trained with Adam (learning rate $10^{-4}$), batch size 16, mixed precision, and up to 500 epochs with early stopping based on the validation loss. We use He-normal initialization and dropout with probability 0.2. Training follows the joint objective defined in Sec.~\ref{sec:training} and is applied on CF channels 0--32 without additional correlation terms. To support reproducibility we fix a global seed for data shuffling, weight initialization, and GPU libraries, configure PyTorch for deterministic kernels, and keep the same seed and hyperparameters for all reported runs.

\subsection{Evaluation Protocol}\label{sec:eval}
Unless stated otherwise, all metrics are computed on CF channels 0--32 and averaged across segments. Let $\mathcal{F}_{\!eval}=\{0,\dots,32\}$, let $T$ be the number of time samples in the \SI{50}{ms} window, and let $\varepsilon>0$ be a small constant for numerical stability.

Normalized mean squared error (NMSE): this metric measures the squared error normalized by the target energy per sample and then averaged across time and CFs,
\begin{equation}
\mathrm{NMSE}
= \frac{1}{|\mathcal{F}_{\!eval}|\,T}
   \sum_{f\in\mathcal{F}_{\!eval}}\sum_{t=1}^{T}
   \frac{\big(N(f,t)-\hat N(f,t)\big)^{2}}{N(f,t)^{2}+\varepsilon^{2}}.
\label{eq:nmse}
\end{equation}

Pearson correlation (per CF):Pearson's correlation coefficient (PCC) $r$,is computed between target and estimate for each CF and then averaged across CFs,
\begin{equation}
\mathrm{PCC}
= \frac{1}{|\mathcal{F}_{\!eval}|}
   \sum_{f\in\mathcal{F}_{\!eval}}
   r\!\big(N(f,:),\hat N(f,:)\big),
\label{eq:corr}
\end{equation}
where $r(\cdot,\cdot)$ denotes the Pearson correlation coefficient of two equal-length sequences.

Signal-to-noise ratio (SNR): a global fidelity measure comparing average signal power to average error power; higher is better.
\begin{equation}
\mathrm{SNR} =
\frac{\tfrac{1}{|\mathcal{F}_{\!eval}|\,T}\sum_{f\in\mathcal{F}_{\!eval}}\sum_{t=1}^{T} N(f,t)^{2}}
{\tfrac{1}{|\mathcal{F}_{\!eval}|\,T}\sum_{f\in\mathcal{F}_{\!eval}}\sum_{t=1}^{T} \big(N(f,t)-\hat N(f,t)\big)^{2}}
\label{eq:snr}
\end{equation}

\subsection{Runtime Measure}\label{sec:runtime}
Streaming inference is evaluated with \SI{150}{ms} of context and a \SI{50}{ms} hop using a batch size of one. For each model we report the latency per frame in milliseconds, defined as the wall-clock time to produce a \SI{50}{ms} output window, the throughput in frames per second, defined as the number of \SI{50}{ms} windows processed per second, and the real-time factor $\mathrm{RTF}=\text{Throughput}\times 0.05$ expressed in audio-seconds per wall-second so that values above one indicate faster-than-real-time processing for a \SI{50}{ms} hop. Benchmarks use \SI{10}{} warm-up iterations and \SI{200}{} timed iterations with explicit CUDA synchronization on an NVIDIA A100-SXM4-80GB in FP16.

\section{Results and Discussion}
All results are reported for CF channels 0--32 (i.e., frequencies smaller than \SI{8}{kHz}; see Sec.~\ref{sec:exp}). We use PCC as the primary metric and additionally report NMSE and SNR, following Sec.~\ref{sec:exp}.

\subsection{PCC by split and CF band}
Table~\ref{tab:pcc_by_split_band} summarizes PCC (mean$\pm$std) per data split (rows) and CF band (columns). 
The overall PCC on the test set, averaged across CF~0--32, is 0.931. 
Also, the average NMSE computed as $10\log_{10}(\mathrm{NMSE})$ is $-10.5$~dB.

\begin{table}[!t]
\centering
\caption{PCC (mean$\pm$std) by data split (rows) and CF band (columns). Overall test PCC over CF 0--32: 0.931.}
\label{tab:pcc_by_split_band}
\setlength{\tabcolsep}{6pt}
\footnotesize
\begin{tabular}{lccc}
\toprule
 & Low (0--16) & Mid (17--28) & High (29--32) \\
\midrule
Train & $0.905 \pm 0.079$ & $0.951 \pm 0.015$ & $0.972 \pm 0.005$ \\
Val   & $0.924 \pm 0.076$ & $0.947 \pm 0.015$ & $0.965 \pm 0.005$ \\
Test  & $0.901 \pm 0.075$ & $0.944 \pm 0.015$ & $0.962 \pm 0.007$ \\
\bottomrule
\end{tabular}
\end{table}

\subsection{Per-CF correlation}
\begin{figure}[!t]
    \centering
    \includegraphics[width=\linewidth]{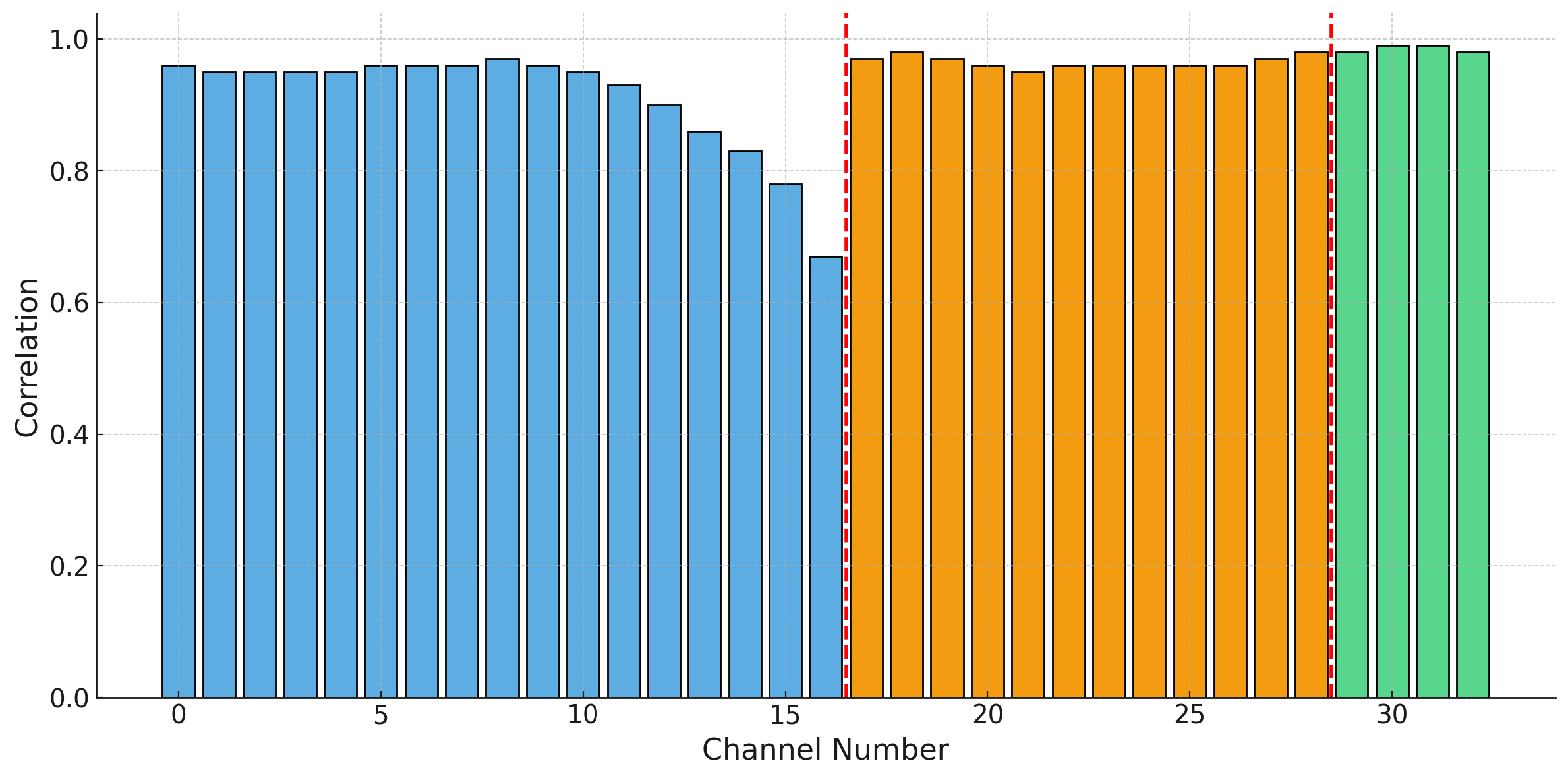}
    \caption{Per-CF PCC on the test set (CF 0--32). Vertical dashed lines indicate band boundaries (0--16, 17--28, 29--32).}
    \label{fig:corr_per_channel}
\end{figure}

Figure~\ref{fig:corr_per_channel} shows PCC per CF on the test set; vertical dashed lines mark the band boundaries. A mild dip appears near the boundary between the low and mid bands (channels 15--16), while PCC generally increases toward higher CFs.

\subsection{SNR by split and CF band}
Table~\ref{tab:db_by_band_split} reports the signal-to-noise ratio, in dB, computed as $10\log_{10}(\mathrm{SNR})$, per split and CF band. 
SNR is slightly lower toward the highest CFs, consistent with the larger dynamic range in that region.

\begin{table}[!t]
\centering
\caption{Signal-to-noise ratio, reported as $10\log_{10}(\mathrm{SNR})$ in dB, by split (rows) and CF band (columns); higher is better.}

\label{tab:db_by_band_split}
\setlength{\tabcolsep}{12pt}
\footnotesize
\begin{tabular}{lccc}
\toprule
 & Low (0--16) & Mid (17--28) & High (29--32) \\
\midrule
Train & $9.9$ & $12.2$ & $13.6$ \\
Val   & $9.6$ & $11.8$ & $12.9$ \\
Test  & $9.4$ & $11.6$ & $12.5$ \\
\bottomrule
\end{tabular}
\end{table}

\subsection{Runtime and real-time performance}
Streaming inference uses \SI{150}{ms} context and a \SI{50}{ms} hop (batch=1). 
As defined in Sec.~\ref{sec:exp}, Table~\ref{tab:rt_compare_icassp} reports latency, throughput, and RTF; 
the model runs faster than real time (RTF $> 2$).

\begin{table}[!t]
\centering
\caption{Per-frame latency and real-time factor on A100--SXM4--80GB (\SI{150}{ms} context, \SI{50}{ms} hop, batch=1).}
\label{tab:rt_compare_icassp}
\setlength{\tabcolsep}{3pt}
\footnotesize
\begin{tabular}{@{}%
  >{\raggedright\arraybackslash}p{0.48\columnwidth}%
  S[table-format=5.2,round-mode=places,round-precision=2]%
  S[table-format=2.3,round-mode=places,round-precision=3]%
  S[table-format=1.5,round-mode=places,round-precision=5,group-digits=false,group-separator={} ]%
@{}}
\toprule
{Model} & {Latency (ms)} & {Throughput (/s)} & {RTF ($\times$)} \\
\midrule
Ours (PyTorch, FP16)                                & 21.54    & 46.420 & 2.32100 \\
Bruce (MATLAB, best-effort GPU)                     & 41310.10 & 0.024  & 0.00121 \\
Bruce (MATLAB, CPU\textsuperscript{$\dagger$})      & 45441.10 & 0.022  & 0.00110 \\
\bottomrule
\end{tabular}
\par\smallskip\noindent\footnotesize$\dagger$ Estimated from the log statement ``GPU is 1.1$\times$ faster''.
\end{table}

\subsection{Discussion}
The encoder attains high overall fidelity with a test-set PCC of 0.931 across CF 0--32 (Table~\ref{tab:pcc_by_split_band}). Correlation increases with CF and shows a mild dip at the low/mid band boundary (Fig.~\ref{fig:corr_per_channel}), likely reflecting differences in dynamic range and time constants across bands. Amplitude fidelity degrades slightly toward higher CFs (Table~\ref{tab:db_by_band_split}), but remains consistent with the trends seen in PCC. Runtime measurements indicate real-time operation by a wide margin (RTF $>$ 2) on a single A100 GPU (Table~\ref{tab:rt_compare_icassp}), supporting streaming use with the \SI{50}{ms} hop.

\section{Conclusion and Future Work}
We introduced a fast, fully differentiable auditory encoder that maps raw audio signals to multi-CF rate-domain neurograms while approximating the deterministic stages of the Bruce model pathway. On clean speech with 33 CF channels, the model attains PCC $\approx 0.93$ and $10\log_{10}(\mathrm{NMSE}) \approx -10.5\,\mathrm{dB}$ with real-time throughput on modern GPUs, providing a practical, reproducible front end for neurogram-based pipelines.

 A small dip appears near the boundary of the low and middle frequency bands, suggesting sensitivity to the band partition; moreover, training and evaluation used clean speech sampled at 16kHz, so robustness to noise, reverberation, and higher-bandwidth sources remains to be quantified.

Future work includes: (i) a deeper study of band transitions (e.g., overlap or shared boundary layers), (ii) evaluation on broader datasets with noise and reverberation and at higher sampling rates, and (iii) development of an inverse decoder from neurograms to audio, including end-to-end training with the encoder. The proposed encoder can also be applied to downstream audio signal processing and auditory neuroscience studies.

% --- סוף גוף המאמר ---
%\startrefsafterfourpages

%\printbibliography

\vfill
\pagebreak

\bibliographystyle{IEEEbib}
\bibliography{references}
\end{document}